\documentclass[twocolumn]{IEEEtran}

\usepackage{amsmath,amssymb,amsfonts,cuted,amsthm}
\usepackage[final]{graphicx}
\usepackage{psfrag}
\usepackage{epsfig}
\usepackage[numbers,sort&compress]{natbib}
\usepackage{array,booktabs}
\usepackage[nolist,nohyperlinks]{acronym}
\usepackage{ucs} 
\usepackage[utf8x]{inputenc}
\usepackage[usenames,dvipsnames]{xcolor}
\usepackage{tikz}
\usepackage{tkz-tab}
\usepackage{multirow}
\usepackage{latexsym}
\usepackage{mathrsfs}
\usepackage{siunitx}
\usepackage{tikz}
\usepackage{tikz-3dplot}
\usepackage{circuitikz}
\usepackage{subfigure}
\usepackage{float}
\usepackage{arydshln}

\usetikzlibrary{calc}

\makeatletter
\def\blfootnote{\xdef\@thefnmark{}\@footnotetext}
\makeatother

\definecolor{bluerevision}{RGB}{0,0,0}

\begin{acronym}
    \acro{5G}{fifth generation}
    \acro{6G}{sixth generation}
    \acro{3GPP}{3rd Generation Partnership Project}
    \acro{BER}{bit error rate}
    \acro{TWDP}{two-wave with diffuse power}
    \acro{FTR}{fluctuating two-ray}
    \acro{MFTR}{multi-cluster fluctuating two-ray}
    \acro{IFTR}{independent fluctuating two-ray}
    \acro{ML}{machine learning}
    \acro{KPI}{key performance indicator}
    \acrodefplural{KPI}[KPIs]{key performance indicators}
    \acro{LoS}{line-of-sight}
    \acro{MIMO}{multiple-inputs multiple-outputs}
    \acro{GTR-V}{generalized two-ray model with von Mises phase distribution}
    \acro{SWAT}{Smart and Wireless Applications and Techonologies}
    \acro{PDF}{probability density function}
    \acrodefplural{PDF}[PDFs]{probability density functions}
    \acro{GDA}{gradient descent algorithm}
    \acro{GA}{genetic algorithm}
    \acro{CDF}{cumulative density function}
    \acrodefplural{CDF}[CDFs]{cumulative density functions}
    \acro{GoF}{goodness of fit}
    \acro{CIR}{channel impulse response}
    
    \acro{nLoS}{non line-of-sight}
    \acro{SNR}{signal-to-noise ratio}
    \acrodefplural{SNR}[SNRs]{signal-to-noise ratios}
    \acro{AWGN}{additive white Gaussian noise}
    \acro{CSI}{channel state information}
    \acro{GMGF}{generalized moment generating function}
    \acro{MGF}{moment generating function}
    \acro{RV}{random variable}
    \acro{MRT}{maximal ratio transmission}
    \acro{MRC}{maximal ratio combining}
    \acro{dR}{double-Rayleigh}
    \acro{CLT}{central limit theorem}
    \acro{i.i.d.}{independent and identically distributed}
    \acro{fdRLoS}{fluctuating double-Rayleigh with line-of-sight}
    \acro{fSOSF}{fluctuating second order scattering fading}
    \acro{dRLoS}{double-Rayleigh with line-of-sight}
    \acro{SOSF}{second order scattering fading}
    \acro{OP}{outage probability}
    \acro{MC}{Monte Carlo}
    \acro{MPC}{multipath component}
    \acrodefplural{MPC}[MPCs]{multipath components}
\end{acronym}

\makeatletter
 \let\old@ps@headings\ps@headings
 \let\old@ps@IEEEtitlepagestyle\ps@IEEEtitlepagestyle
 \def\confheader#1{%
 \def\ps@headings{%
 \old@ps@headings%
 \def\@oddhead{\strut\hfill#1\hfill\strut}%
 \def\@evenhead{\strut\hfill#1\hfill\strut}%
 }%
 \def\ps@IEEEtitlepagestyle{%
 \old@ps@IEEEtitlepagestyle%
 \def\@oddhead{\strut\hfill#1\hfill\strut}%
 \def\@evenhead{\strut\hfill#1\hfill\strut}%
 }%
 \ps@headings%
 }
 \makeatother
\confheader{%
This work has been accepted for publication in \textit{IEEE Transactions on Wireless Communications}. DOI: 10.1109/TWC.2024.3509738
}

\begin{document}
\title{Empirical Validation of a Class \\ of Ray-Based Fading Models}

\author{Juan E. Galeote-Cazorla, Alejandro Ramírez-Arroyo, F. Javier Lopez-Martinez and Juan F. Valenzuela-Valdés}


\maketitle

\begin{abstract}
    As new wireless standards are developed, the use of higher operation frequencies comes in hand with new use cases and propagation effects that differ from the well-established state of the art. Numerous stochastic fading models have recently emerged under the umbrella of \textit{generalized fading} conditions to provide a fine-grain characterization of propagation channels in the mmWave and sub-THz bands. For the first time in literature, this work carries out an experimental validation of a class of such ray-based models in a wide range of propagation conditions~(anechoic, reverberation and indoor scenarios) at mmWave bands. These models allow to characterize the communication channel with a reduced number of physically interpretable parameters. In specific, we show that the independent fluctuating two-ray~(IFTR) model has good capabilities to recreate rather dissimilar environments with high accuracy and only four parameters. We also put forth that the key limitations of the IFTR model arise in the presence of reduced diffuse propagation, and also due to a limited phase variability for the dominant specular~components. 
\end{abstract}
    

\blfootnote{\noindent This work has been supported by grant TED2021-129938B-I00 funded by MCIN/AEI/10.13039/501100011033 and by the European Union NextGenerationEU/PRTR. It has also been supported by grants PID2020-112545RB-C54, PDC2022-133900-I00, PDC2023-145862-I00 and TED2021-131699B-I00, funded by MCIN/AEI/10.13039/501100011033 and by the European Union NextGenerationEU/PRTR; in part by Junta de Andaluc\'ia through grant EMERGIA20-00297; and in part by the predoctoral grant FPU22/03392 (\textit{Corresponding author: Juan E. Galeote-Cazorla}).}
    
\blfootnote{\noindent Juan E. Galeote-Cazorla, F. Javier Lopez-Martinez and Juan F. Valenzuela-Valdés are with the Department of Signal Theory, Telematics and Communications, Research Centre for Information and Communication Technologies (CITIC-UGR), University of Granada, 18071, Granada, Spain. \linebreak F. Javier Lopez-Martinez is also with the Communications and Signal Processing Lab, Telecommunication Research Institute (TELMA), Universidad de M{\'a}laga, M{\'a}laga, 29010, Spain (e-mails: juane@ugr.es; fjlm@ugr.es; \linebreak juanvalenzuela@ugr.es)}

\blfootnote{\noindent Alejandro Ramírez-Arroyo is with the Department of Electronic Systems, Aalborg University (AAU), 9220 Aalborg, Denmark (e-mail:araar@es.aau.dk).}

\section{Introduction}\label{sec:Introduction}
\IEEEPARstart{T}{}he advent and widespread deployment of the \ac{5G} of wireless technology has been key to improve the achievable data rates well-beyond tenths of Gbps, and also to reduce latency in one order of magnitude with respect to previous standards --- now approaching to 1\:ms~\cite{Rappaport2017_5G_overview}. One of the key enablers for such achievements is the use of higher frequency bands in the millimeter waves (mmWave) range. This trend is likely to continue as \ac{5G} evolves, and sixth generation~(6G) is often envisioned to provide coverage in the sub-THz range for some use cases \cite{Polese2020}.

To properly modeling and describing propagation channels, different approaches can be taken to capture the true (and often intractable) nature of electromagnetic effects as signals traverse the wireless environment. Classically, channel modeling strategies have been categorized as empirical vs. analytical, and also as deterministic vs. stochastic \cite{Matolak2008}. Today, state-of-the-art channel models combine the key features from these approaches to improve accuracy. For instance, COST \cite{cost1999_cost231} and 3GPP-like \cite{rangos_5G} models combine empirical measurements (e.g. for path-loss exponents) with geometric modeling of the propagation environment, also incorporating some stochastic features to provide randomness inherent to wireless fading effects. These models often have hundreds of parameters that allow to recreate wireless propagation with high versatility, at the expense of a higher complexity and computational burden. Recently, ray-tracing approaches have been considered as an alternative to provide a realistic channel characterization in terms of a number of \acp{MPC} \cite{Lecci2021}, at the price of an overwhelming mathematical complexity. 

To circumvent the intricacy of these approaches, analytical stochastic channel models are of widespread use due to their comparatively reduced complexity, much fewer parameters, and mathematical tractability \cite{Simon2004_rician_rayleigh}. Rayleigh and Rician models are immensely popular in the wireless community, as the de facto standards to model small-scale fading in non line-of-sight~(nLoS) and \ac{LoS} cases, respectively. As new use cases promote the ubiquity of wireless devices, many new environments in sensor networks and industrial settings require the development of more advanced models\footnote{While any distribution borrowed from statistics may be used to approximate the behavior of wireless channels, only a certain class of distributions that comply with electromagnetic propagation laws will capture the true nature of communications channels.}, often referred to as \textit{generalized}. For instance, the \ac{TWDP} model \cite{Durgin2002_TWDP} proposed by Durgin, Rappaport and de Wolf describes a multipath environment with two dominant constant-amplitude specular rays, plus an aggregate diffuse component. With only one parameter addition compared to the Rician case, the \ac{TWDP} model has an increased capability to recreate not-so-common propagation conditions; these include bimodality and hyper-Rayleigh behavior \cite{Frolik2008}. Subsequently, different generalizations and alternatives have been proposed in the literature: for instance, the \ac{FTR} model \cite{RomeroJerez2017_FTR_model} allows that the two rays of the \ac{TWDP} model fluctuate \textit{jointly} or \textit{independently} \cite{Olyaee2022_IFTR}. This philosophy is inherited from Abdi's Rician-shadowed model~\cite{Abdi2003_Rician_Shadowed}, which also generalizes the Rician one. The \linebreak $\kappa$-$\mu$ and the $\eta$-$\mu$ models proposed by Yacoub add additional flexibility to the Rician and Rayleigh models through the notion of multipath wave clusters. These were also later generalized by Paris' $\kappa$-$\mu$ shadowed fading model \cite{Paris2014}. The goal of ray-based models is to improve the fitting performance in real environments avoiding the aforementioned use of a large number of parameters with unclear physical meaning.

Recently, it was shown that it is possible to bridge the gap between the models used for industry (e.g., 3GPP TR 38.901) and academia (e.g., ray based), by developing a \ac{MIMO} channel model that calibrates the few parameters of the latter using the full 3GPP TR 38.901 channel model as a reference \cite{Pagin2023}. Clearly, the key to use any of the aforementioned statistical models for \ac{5G} performance evaluation in practical conditions is their empirical validation over a great variety of scenarios. One way to accomplish this is through measurement campaigns on a specific scenario; for instance, mobile radio channels over the sea at 5.9\:GHz or vehicular channels during overtakes at 60\:GHz exhibit \ac{TWDP} behaviors \cite{Zochmann2019_vehicular_TWDP,Yang2018_sea_TWDP}. However, such campaigns are rather costly in terms of time and resources, and cannot be replicated in a controlled way. Another alternative is the use of laboratory chambers combining anechoic and reverberation features. These allow to recreate Rician-fading~\cite{Holloway2006_Rician_AnechoicChamber} or two-ray~\cite{Frolik2009} environments, which can be exploited to emulate \ac{MIMO} channels with such particular behavior \cite{SanchezHeredia2011_RicianMIMO}.

In this paper, we aim to empirically validate\footnote{In this context, this terminology is used in a broader sense than in experimental physics literature, where the classical approach for experimental validation is hypothesis testing. However, the state of the art in channel modeling often discards it in favor of a model-based one \cite{ref:Molisch2005}. That is, empirical validation refers to the process to properly fit and describe measurements with theoretical models previously proposed in the literature. This is the approach we consider here for empirical validation, and the one followed in other works~as~\cite{ref:Kermoal2002}.} a class of ray-based stochastic fading models in a wide variety of scenarios. Specifically, we used our mixed anechoic/reverberation chamber facilities, together with indoor acquisition, to find a concordance between measurements and different propagation models in the frequency band from 24.25\:{GHz} to 27.5\:GHz (i.e., 3GPP n258 band), which belongs to the mmWave general classification \cite{3GPP2022_TR38.901}. For this purpose, we consider the independent fluctuating two-ray (IFTR) fading model recently proposed in [11], which seemed more adequate than the original FTR model \cite{RomeroJerez2017_FTR_model} or further generalizations of it for this specific context. The IFTR model offers a great flexibility to model the wide spectrum of experimental channels through only four physically interpretable parameters. Finally, this model showed a remarkable \ac{GoF} in most scenarios. Nevertheless, we also identified certain scenarios which were not properly fitted with the IFTR. Interestingly, we identify that the widespread assumption of uniform phases for the two dominant specular components \cite{Durgin2002_TWDP} does not hold in some scenarios with reduced multipath. We show that in those scenarios where this behavior is identified, a more general modeling for the phases \cite{Rao2015_GTR-V} improves the fitting performance. In brief, we successfully validate these ray-based models in a wide spectrum of environments and rather dissimilar conditions with a detailed analysis of their physically meaningful~parameters.

The remainder of the article is organized as follows. Section~\ref{sec:IFTR} summarizes the \ac{IFTR} propagation model with the relevant physical and mathematical basis. Section \ref{sec:Setup} explains how the measurements were acquired and how they are used for recreating new scenarios. Section \ref{sec:Results} shows and discuss the obtained results from fitting the scenarios with the propagation models. Finally, Section \ref{sec:Conclusions} summarizes the key conclusions and the future lines derived from this work.

\section{Physical Channel Model}\label{sec:IFTR}
Let us consider a general formulation for the received radio signal over a wireless channel~\cite{durgin2000theory}, given as a superposition of a number of waves
\begin{align}
\label{eq1}  
V_r =\sum_{n=1}^{M} A_n \exp\left({j\varphi_n}\right),
\end{align}
where $A_n$ and $\varphi_n$ denote their corresponding amplitudes and phases. Now, it is possible to rewrite \eqref{eq1} in the following way:
\begin{align}
\label{eq2}  
V_r =\sum_{n=1}^{N} A_n \exp\left({j\varphi_n}\right) + \underbrace{\sum_{m=1}^{P} A_m \exp\left({j\varphi_m}\right)}_{Z},
\end{align}
with $M=N+P$, so that $N$ now represents a group of dominant specular waves, while $P$ indicates a group of numerous and relatively weak diffusely propagating waves. For sufficiently large $P$, the \ac{CLT} holds for the second term in \eqref{eq2}, implying that $Z\triangleq\sum_{m=1}^{P} A_m \exp\left({j\varphi_m}\right)$ can be approximated as a complex Gaussian random variable~(RV) with zero-mean and variance $2\sigma^2$. For $N$\:=\:2, constant-amplitude $A_n$ and uniformly distributed phases $\varphi_n$, the \ac{TWDP} model emerges \cite{Durgin2002_TWDP}. In the sequel, and taking the TWDP model as a baseline reference, we consider the general case on which $A_n=V_n\sqrt{\xi_n}$ so that $V_n$ denotes the amplitude of each of these dominant specular components, whereas~$\xi_n$ are unit-mean independent Gamma RVs characterizing amplitude fluctuations. This model, referred to as IFTR model, was recently formulated in \cite{Olyaee2022_IFTR}, and is fully characterized by the following set of parameters:
\begin{align}
	K &= \frac{V_1^2+V_2^2}{2\sigma^2} \in [0,\infty),\\
	\Delta &= \frac{2V_1V_2}{V_1^2+V_2^2} \in [0,1],
\end{align}
\begin{align}
    m_n\,&\{n=1,2\} \in (0,\infty),\\
    \Omega &\triangleq \rm E\{|V_r|^2\} \in [0,\infty),
\end{align}
where $1/m_n$ denotes the fading severity (i.e., the amount of fluctuation) for each of the specular components, $K$ is the Rician factor defined as the ratio between the average powers of the dominant specular waves and the diffuse components, and $\Delta$ captures the amplitude dissimilarity between the two dominant specular waves. Finally, the scale parameter $\Omega=~V_1^2+V_2^2+2\sigma^2$ represents the average received power.

\begin{figure*}
    \begin{multline}\label{eq:IFTR-pdf}
        f_r(r) = \frac{2r\left(1 + K\right)}{\Omega}m_1^{m_1}m_2^{m_2}\left(m_1 + \frac{K}{2}\left(1 + \sqrt{1 - \Delta^2}\right)\right)^{m_2 - m_1}\sum_{n = 0}^{m_1 - 1}\frac{1}{n!}\binom{m_1 - 1}{n}\frac{\Gamma(m_2 + n)}{\Gamma(m_2)}\left(\frac{K\Delta}{2}\right)^{2n} \\
        \times\left[m_1\frac{K}{2}\left(1 - \sqrt{1 - \Delta^2}\right) + m_2\frac{K}{2}\left(1 + \sqrt{1 - \Delta^2}\right) + m_1m_2\right]^{-m_2-n} \\
        \times\Phi_2^{(3)}\left(n + 1 - m_1,m_1 - m_2,m_2 + n;1;-\frac{1 + K}{\Omega}r^2,-\frac{m_1\left(1 + K\right)}{\left(m_1 + K/2\left(1 + \sqrt{1 - \Delta^2}\right)\right)\Omega}r^2, \right. \\
        \left. -\frac{m_1m_2\left(1 + K\right)}{\left(m_1\:K/2\left(1 - \sqrt{1 - \Delta^2}\right) + m_2\:K/2\left(1 + \sqrt{1 - \Delta^2}\right) + m_1m_2\right)\Omega}r^2\right).
    \end{multline}
    \rule{\linewidth}{0.5pt}
\end{figure*}

{A closed-form expression for the \ac{PDF} $f_r(r)$ of the received radio signal magnitude $r = |V_r|$ can be found in (\ref{eq:IFTR-pdf}). It is expressed in terms of the parameters defined in eqs. (3)-(6) and the confluent hypergeometric function~$\Phi_2^{(n)}(\cdot)$ \cite{phi_function}. This expression is only valid when $m_1$ and $m_2$ take integer values. Nevertheless, every positive real value can be evaluated by computing the numerical inverse Laplace transform over the \ac{MGF} \cite{Olyaee2022_IFTR}. Moreover, a new analytical PDF based on mixture of Gamma distributions has been recently \linebreak proposed in \cite{Olyaee2022_PDF_IFTR}.}

The main advantage of these ray-based statistical models is their physical interpretability and the reduced number of parameters. Alternatives such as the mixture of distributions (e.g. Gaussian or Gamma mixture models~\cite{Selim2016},~\cite{Papasotiriou2023}) can fit experimental data with arbitrary precision. However, they lack any physical interpretation as the number of involved distributions increases. In contrast, models such as the IFTR only need a reduced set of parameters with a clear physical meaning as previously discussed. They allow us to characterize channels in terms of their physical behavior.

\section{Measurements Description}\label{sec:Setup}
The studied measurement dataset consists in the channel frequency response obtained for three different scenarios, namely: semi-anechoic and semi-reverberation chambers, and indoor. A more extensive description of the channel environments can be consulted in \cite{RamirezArroyo2022_tSNE}. They were acquired at the Smart and Wireless Applications and Technologies~(SWAT) research group facilities, located at the University of Granada~(Spain). Acquisition was performed with a vector network analyzer (VNA Rohde \& Schwarz ZVA67) as the channel sounder, which is capable of measuring the S-parameters up to 67\:GHz. The analyzed frequency band sweeps from 24.25\:GHz to 27.5\:GHz with a 5\:MHz step, which leads to 651 measured frequency points. It corresponds with the 3GPP n258 band, included in the mmWave classification. Standardized gain horns were used as transmitter~(Tx) and receiver (Rx) antennas, specifically the Flann K-band antenna model \#21240-20 fed with a WR-34 waveguide-to-coaxial transition. Finally, the transmission power was set to~10\:dBm.

\begin{figure}[t]
    \centering
    \subfigure[Semi-anechoic and semi-reverberation chamber.]{\includegraphics{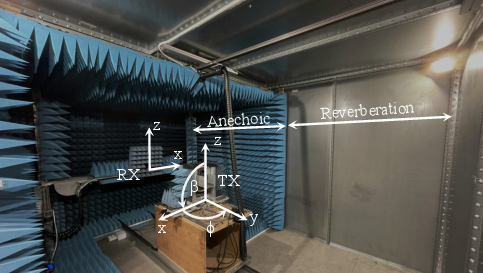}\label{fig:semianechoic_chamber_scheme}}
    \par\bigskip
    \subfigure[Indoor laboratory.]{\includegraphics{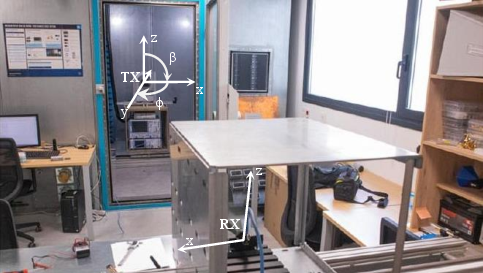}\label{fig:indoor_setup}}
    \caption{Panoramic picture of (a) the chamber environments, and (b) the indoor scenario. The receiver antenna can move over the entire XZ plane. However, the transmitter only modifies the $\phi$ and $\beta$ angles, named azimuth and roll respectively, for changing orientation and polarization.}
\end{figure}

As previously mentioned, two of the scenarios were measured in a semi-anechoic semi-reverberation chamber whose dimensions are 5$\times$3.5$\times$3.5 meters [see Fig. \ref{fig:semianechoic_chamber_scheme}]. The semi-anechoic side is completely covered with absorbents to suppress any possible reflection in order to receive the ray from the \ac{LoS} path exclusively. In contrast, the semi-reverberation side is composed entirely of metallic walls favoring the appearance of multiple reflections. The emerged behavior corresponds with a multipath channel. In contrast, the indoor scenario is slightly different. It corresponds with the experimental laboratory of the research group, where the chamber is located. In this case, the scenario represents by itself a real environment for communications, whereas the measurements in chambers can be used for the emulation of new channels.

The measuring setup for the semi-anechoic chamber consists of both Tx and Rx in \ac{LoS} configuration with a distance between each other of 160\:cm. The~Rx antenna describes an 11$\times$11 virtual uniform rectangular array (URA) over the XZ plane with a constant separation of 4\:cm between the elements. On the other hand, the Tx antenna sweeps three azimuth angles $\phi$: 0º and $\pm$\:30º, where the angle 0º corresponds with the perpendicular line pointing to the XZ plane. In addition, three roll angles $\beta$ are swept: 0º and~$\pm$\:30º, where now the angle 0º represents the situation with no-depolarization losses. This implies 3$\times$3 different configurations for the Tx antenna and a total number of point-to-point measured channels of 11$\times$11$\times$3$\times$3\:=\:1089. These propagation measurements are characterized by a single ray and a reduced diffuse power due to the absorbents.

In the semi-reverberation chamber, the setup changes to a nLoS situation where the Tx antenna is pointing to the reverberation side of the chamber [see Fig. \ref{fig:semianechoic_chamber_scheme}]. Now, the distance between the antennas is 600\:cm, measured as the sum of the distances of each antenna to the metallic wall parallel to the XZ plane. The Tx and Rx antenna configurations are analogous to those described for the anechoic chamber, obtaining additional 1089 channel measurements. For this case, the azimuth angle $\phi$\:=\:0º represents the situation where the Tx antenna points to the metallic wall parallel to the XZ plane. In this scenario, multipath channels are generated where up to seven reflections are identified in our experiments depending on the Tx-Rx configuration. Since it is a semi-reverberation chamber, the measurements involve several individual rays instead of a continuous reception as one would expect in a full reverberation chamber. This is a consequence of the absorption at the anechoic side. Each ray has a particular power that can be significantly higher than the noise floor or comparable to it, classifying them as specular or diffuse components respectively.

Finally, in the indoor setup the Tx antenna is located inside the chamber pointing outside to the Rx antenna in LoS, which is positioned inside the laboratory room [see Fig.~\ref{fig:indoor_setup}]. It is furnished with several desks, chairs and shelves. The walls are made of concrete and have metal-edge windows. About the setup configuration, the Rx describes the same 11\:$\times$\:11 virtual URA as for the chamber scenarios with a separation between elements of~4\:cm. However, the Tx changes the azimuth angle~$\phi$, now sweeping $-$30º, $-$15º, and 0º for exclusively pointing outside the chamber in a \ac{LoS} situation across the chamber right side window and the door. The swept~$\beta$ angles remain the same, that is, $\pm$30º and~0º. Thus, the total number of measured channels for the indoor scenario is also~1089. As a complex environment, several specular and diffuse components are received along with the \ac{LoS} ray.

A TOSM (Through -- Open -- Short -- Match) calibration process has been performed, setting the calibration planes at the end of the coaxial lines that connect the VNA with the horn antennas. Thus, all the acquired measurements include the antenna gain patterns as components of the propagation channel. Then, we can admit that the channel response $H(f)$ is equivalent to the measured transmission parameter $S_{21}(f)$. On the other hand, the received signal $Y(f)$ from a linear and time-invariant system can be computed as $Y(f)$\:=\:$H(f)\:X(f)$, where $X(f)$ is the transmitted signal. If we assume that $X(f)$\:=\:1\:V, we can admit that all the VNA measurements are expressed in terms of volts. Additionally, each of the 651 frequency points can be considered as an individual realization of the channels because of the static conditions of the environments during the measurement process.

\begin{figure}[t]
    \centering
    \subfigure[{Selection of two different positions in the receiver array.}]{\includegraphics{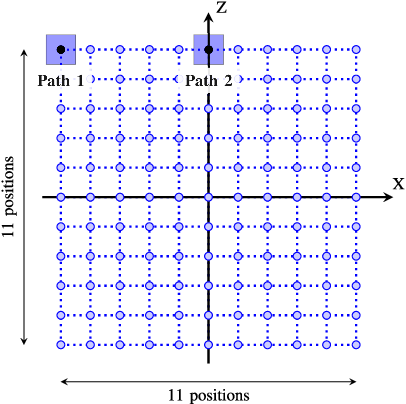}\label{fig:receiver_system}}
    \par\bigskip
    \subfigure[{Transmitter orientation of 0\:º in azimuth and roll angles.}]{\includegraphics{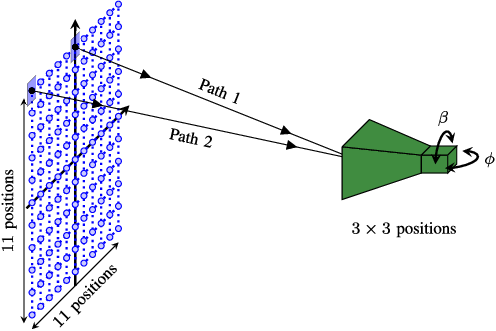}\label{fig:2-ray_selection_in_receiver}}
    \caption{{Channel combination example (configuration A) of two receiver positions linked to the same transmitter orientation.}}
    \label{fig:complete_channel_scheme_in_reverse_path}
\end{figure}

The described setups have resulted in 1089 channel measurements per scenario\footnote{We are focusing on point-to-point measurements. However, this setup might be used for studying MIMO systems by synthesizing arbitrary planar array geometries \cite{Wilding2023}. Moreover, performance comparisons between distributed and centralized MIMO might be carried out.} (anechoic, reverberation and indoor). By construction, the chamber scenarios may not resemble propagation channels with more than one specular dominant component for communications. Nevertheless, more complex scenarios can emerge from the combination of the original measurements. Note that similar approaches are carried out in the literature to generate synthetic array geometries from individual measurements \cite{Wilding2023}. A simple application of this methodology is followed for this work. If we denote $H_1(f)$ and $H_2(f)$ as two different channel measurements from the same original scenario, we can compute a more complex channel $\hat{H}(f)$ by performing an equally-weighted linear combination as follows:
{\begin{equation}\label{eq:linear_combination}
    \hat{H}(f) = H_1(f) + H_2(f).
\end{equation}}
In the anechoic case a low diffuse power and the presence of only one relevant ray are expected. When two measurements are combined by applying \eqref{eq:linear_combination}, the resulting channel may reproduce scenarios where the diffuse power is reduced and only one or two rays are received in a LoS situation, emulating the two-ray scenario conditions intrinsic to the IFTR model. This is closely related to the appearance of bimodality behaviors in the channel PDFs, an aspect that cannot be described with the classical Rician and Rayleigh models. On the other hand, the reverberation case has a strong diffuse power with the presence of multiple rays. The combination of channels may resemble an industrial environment where nLoS situations are common and reflections over metallic surfaces are the main propagation mechanisms. Additionally, linear combinations may also be performed for the indoor measurements. In this case, the advantage relies on the increase of channel diversity since it is equivalent to introducing new reflections or additional diffuse power sources. The fitting of the \ac{IFTR} model with these measurements allow us to validate it in realistic scenarios.

For giving physical coherence to the linear combinations, the channels $H_1(f)$ and $H_2(f)$ have to share the same transmitter orientation or receiver position. That is, if we combine two channels with different Rx array positions, the corresponding Tx orientations must be the same. Similarly, if we take two channels with different Tx orientations, they must be linked to the same Rx array position. Fig. \ref{fig:complete_channel_scheme_in_reverse_path} shows one combined or merged channel example under this criterion, named configuration A in the following. Two positions in the Rx array are selected as in Fig. \ref{fig:receiver_system}. Then, one Tx orientation is fixed for both positions as in Fig. \ref{fig:2-ray_selection_in_receiver}.

The number of possible pairs of linear combinations becomes unmanageable since it is in the order of hundreds of thousands\footnote{Two animations have been included as supplementary material for the reader showing numerous configurations examples and their amplitude \ac{PDF} for the three studied scenarios. They are available in Appendix B.}. This fact implies that only a reduced sample of all the possible combinations shall be selected \cite{sampling}, searching for a compromise between the number of configurations and the total execution time of the IFTR model validation process. Specifically, 142 different configurations have been randomly chosen (they will be the same ones for the three scenarios). Among them, 131 correspond to cases where the combined pairs share a common Tx orientation, and 11 to pairs sharing a common Rx array position. To avoid possible biases, each sub-set has been previously analyzed. In the first case, it has been checked that the distribution of distances between the positions of the selected pairs is approximately equal to the distribution of all the possible pairs in the Rx array. On the other hand, combinations of different Tx orientations lead to three situations: same polarization but different pointing, same pointing but different polarization, and different pointing and polarization. It has been checked that the random sub-set includes combinations of the three types. For a fair comparison, the 142 selected configurations are the same for anechoic, reverberation, and indoor scenarios. In this last case, as previously mentioned, the azimuth angle swept $-$30º, $-$15º, and 0º instead of $-$30º, 0º, and 30º. This implies that configurations of anechoic and reverberation scenarios cannot be replicated exactly for the indoor one. However, by tagging the angles as $-$1, 0, and 1 respectively in both cases, combinations can be computed while keeping the same tag for the three scenarios.

\begin{figure}[t]
    \centering
    \includegraphics{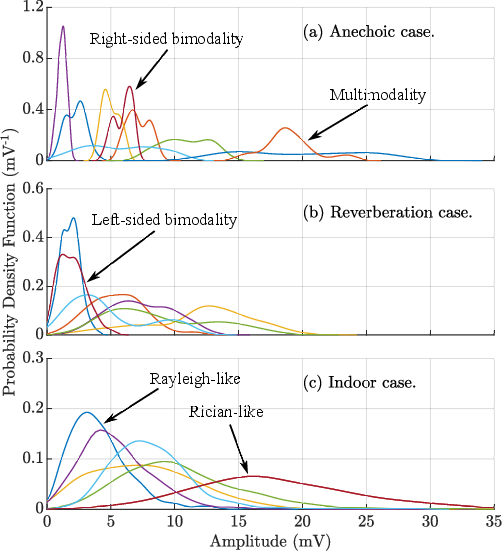}
    \caption{Amplitude \acp{PDF} associated with some examples of merged channels for the three scenarios. Each color represents one individual channel configuration. A great diversity of behaviors is observed (e.g. Rician-like and Rayleigh-like, multimodality...).}
    \label{fig:scenarios_diversity}
\end{figure}

\begin{figure*}[t]
    \centering
    \subfigure[]{
        \centering
        \includegraphics{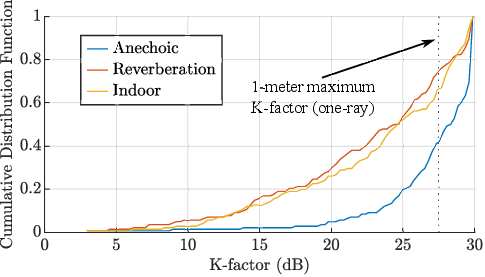}\label{fig:Kfactor}

    }
    \hspace{0.045\linewidth}
    \subfigure[]{
        \centering
        \setcounter{subfigure}{3}   
        \includegraphics{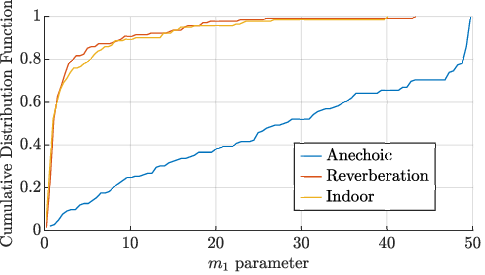}\label{fig:m1}
    }
    \par
    \subfigure[]{
        \centering
        \setcounter{subfigure}{2}
        \includegraphics{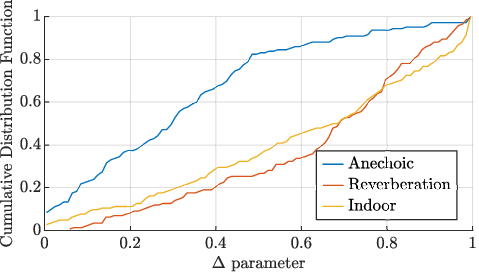}\label{fig:D}
    }
    \hspace{0.045\linewidth}
    \subfigure[]{
        \centering
        \setcounter{subfigure}{4}
        \includegraphics{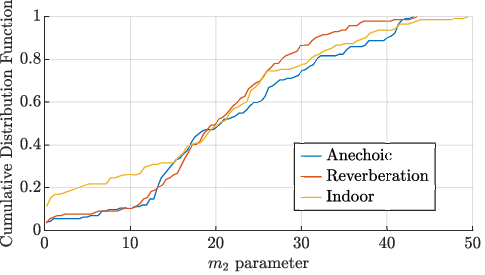}\label{fig:m2}
    }
    \caption{Cumulative distributions of the obtained \ac{IFTR} parameters for the three analysed environments.}
    \label{fig:iftr_parameters_cdf}
\end{figure*}

As a result of performing the linear combinations, a great diversity of channels emerges. Fig. \ref{fig:scenarios_diversity} shows some examples of amplitude PDFs for the three scenarios. A multitude of behaviors such as Rician-like, Rayleigh-like or multimodality are observed. These latter are especially common in the anechoic configurations\footnote{While the merged channels in anechoic case should correspond to a two-ray case by construction, in situations with very reduced multipath the \ac{CLT} does not apply; hence, additional rays [e.g., those arising from reflections in the posts, see Fig. 1(a)] need to be accounted for individually. In this circumstance, the effect of these rays is translated into some additional modality \cite{Romero2022}.}, but also monomodal ones. Bimodality is a particular behavior where two local maxima appear. Ideally, they result from the presence of two rays whose power is substantially higher than the diffuse one or the noise floor \cite{Durgin2002_TWDP}. In the case of our experimental measurements, the merged channels are computed to guarantee the presence of (at least) two specular components arriving at the receiver. 

As mentioned above, a wide spectrum of communication environments are obtained. From a more general point of view, these merged channels might recreate scenarios such as offices, meeting rooms or hallways. The presence of specular and diffuse components would depend on site-specific characteristics (e.g. building materials or obstacles) and the operation frequency, finally resulting in monomodal or multimodal propagation channels. For example, if the LoS component arrives to the receiver together with a strong reflection from a metallic surface, a bimodal channel may naturally emerge.

Frequency is also key to the expected environments behavior. Sub-6\:GHz bands propitiate the appearance of a great number of diffuse components due to the basic propagation mechanisms: reflection, diffraction and scattering. This leads naturally to the classical Rician/Rayleigh scenarios depending on the presence/absence of the LoS component \cite{ref:Andersen1995}. In contrast, for higher frequencies in the sub-THz and THz bands the propagation mechanisms might be different. Diffuse power decreases because of the higher losses and the detriment of some mechanisms such as reflection or diffraction. Then, several individual contributions may stand out from the rest (not only the LoS) naturally emerging multimodal and more complex behaviors \cite{ref:Rappaport2019}. Our particular case, the mmWaves band, is located in between the sub-6\:GHz and the THz bands. Depending on the specific characteristic of the propagation scenarios, channels might be closer to the former or the latter case. Therefore, the great diversity of the merged channels exemplified with Fig.~3 becomes an excellent opportunity to validate the ray-based models at the n258 mmWaves band.

\section{Empirical Validation and Analysis}\label{sec:Results}
\subsection{Analysis of the obtained IFTR parameters}
Once the Tx-Rx configurations have been selected, a fitting between the empirical \acp{PDF} and the \ac{IFTR} theoretical one is carried out. In this work, we have decided to use a methodology based on multi-objective optimization with a \ac{GA}. For a detailed explanation and certain particularities on the fitting methodology, see Appendix A. The \acp{CDF} of the obtained \ac{IFTR} parameters for each scenario are represented in Fig.~\ref{fig:iftr_parameters_cdf}. A high contrast between anechoic and multipath\footnote{The term ``multipath'' refers, henceforth, to both reverberation and indoor propagation scenarios.} scenarios is observed, with clear different trends. 

\begin{table*}[t]
    \centering
    \caption{Numerical results obtained for the given fitting examples in the three studied scenarios}
    \renewcommand{\arraystretch}{1.75}
    \begin{tabular}{c|c|cccccc|}
        \multicolumn{1}{c}{} & \multicolumn{1}{c}{} & \multicolumn{6}{c}{\textbf{Configuration}} \\ \cline{3-8}
        \multicolumn{1}{c}{\textbf{Parameter}} & \multicolumn{1}{c|}{\textbf{Scenario}} & \textbf{A} & \textbf{B} & \textbf{C} & \textbf{D} & \textbf{E} & \textbf{F} \\ \hline
        \multirow{3}{*}{$K$ [\si{\decibel}]} & Anechoic & \num{29.3} & \num{27.8} & \num{29.9} & \num{25.0} & \num{29.9} & \num{30.0} \\
        & Reverberation & \num{25.8} & \num{27.3} & \num{29.4} & \num{29.8} & \num{22.7} & \num{26.3} \\ 
        & Indoor & \num{22.8} & \num{25.6} & \num{29.9} & \num{27.1} & \num{29.9} & \num{20.6} \\ \hline
        \multirow{3}{*}{$\Delta$} & Anechoic & \num{0.44} & \num{0.09} & \num{0.23} & \num{1.00} & \num{0.45} & \num{0.21} \\
        & Reverberation & \num{0.09} & \num{0.80} & \num{0.49} & \num{0.73} & \num{0.33} & \num{0.52} \\
        & Indoor & \num{0.58} & \num{0.24} & \num{0.02} & \num{1.00} & \num{0.54} & \num{0.99} \\ \hline
        \multirow{3}{*}{$(m_1,m_2)$} & Anechoic & $(\num{35},\num{16})$ & $(\num{27},\num{13})$ & $(\num{50},\num{35})$ & $(\num{11},\num{22})$ & $(\num{36},\num{12})$ & $(\num{50},\num{14})$ \\
        & Reverberation & $(\num{12},\num{19})$ & $(\num{1},\num{18})$ & $(\num{1},\num{33})$ & $(\num{0.60},\num{0.39})$ & $(\num{15},\num{27})$ & $(\num{3},\num{32})$ \\
        & Indoor & $(\num{0.69},\num{19})$ & $(\num{0.93},\num{22})$ & $(\num{0.46},\num{15})$ & $(\num{5},\num{38})$ & $(\num{1},\num{25})$ & $(\num{2},\num{25})$ \\ \hline\hline
        \multirow{3}{*}{$\rm RMSE$} & Anechoic & \num{4.10} & \num{16.46} & \num{26.51} & \num{33.48} & \num{9.10} & \num{16.64} \\
        & Reverberation & \num{4.00} & \num{9.71} & \num{4.22} & \num{10.41} & \num{3.10} & \num{2.46} \\
        & Indoor & \num{4.30} & \num{3.52} & \num{8.86} & \num{4.24} & \num{1.10} & \num{3.29} \\ \hline
    \end{tabular}
    \label{tab:fitting_results}
\end{table*}

\begin{figure}[t]
    \centering
    \includegraphics{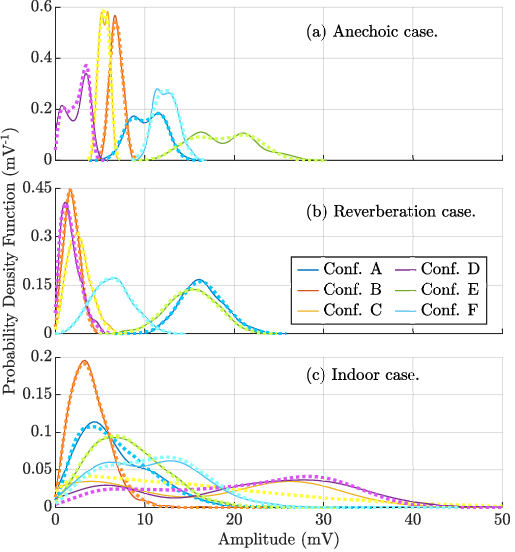}
    \caption{Fitting examples for different configurations in the three analysed scenarios. Solid and dotted lines represent empirical and \ac{IFTR} model distributions respectively.}
    \label{fig:fitting_solutions}
\end{figure}

\begin{figure}[t]
    \centering
    \includegraphics{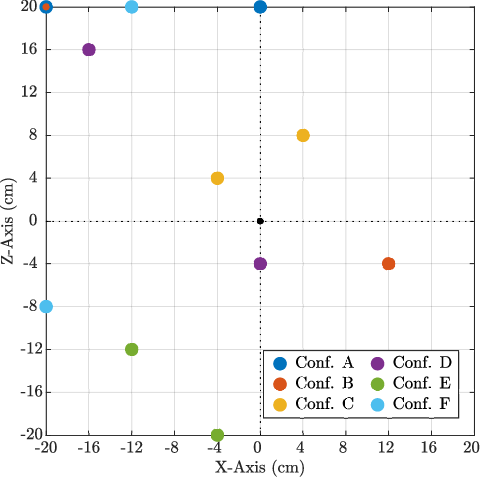}
    \caption{Configurations of the given fitted channel examples. Each pair of points with the same color identifies the two Rx array positions that compose the merged channel. The common Tx orientations (as tags) for each configuration are: $\{\phi,\beta\}$\:=\:$\{$0,0$\}$ for A, E, and F; $\{\phi,\beta\}$\:=\:$\{$-1,0$\}$ for B; and $\{\phi,\beta\}$\:=\:$\{$1,1$\}$ for C and D.}
    \label{fig:configurations}
\end{figure}

From Fig. \ref{fig:Kfactor}, it can be seen that the $K$-factor takes higher values on the anechoic configurations, which is due to the absence of diffuse power. In fact, the median value for this scenario is 28.3\:dB, whereas for the multipath ones is 24.7\:dB; that is, 3.6\:dB of difference. The graphic also shows the maximum measurable $K$-factor for an individual ray at a distance of 1 meter for the given VNA setup. It has been empirically determined in our anechoic chamber, and takes values up to 27.5\:dB. The limit for two-rays would be 3\:dB higher (at most), and since the distances between Tx and Rx in our measurements are significantly larger than 1\:m, we can ensure that there is no merged channel which might surpass an IFTR $K$-factor of 30\:dB as confirmed in Fig. \ref{fig:Kfactor}.

On the other hand, as can be observed in Fig. 4(b), the $\Delta$ parameter in the anechoic case distributes along lower values than for the multipath configurations. That is a consequence of the lack of reflections in the anechoic scenario, where only the two specular rays are relevant and their amplitudes are significantly different. Moreover, it is also remarkable that anechoic and reverberation scenarios are the inverse of each other. For the latter, the multipath specular components that arrive at the antenna have similar amplitudes. That, in addition to the nLoS situation, imposes $\Delta$ taking higher values. The obtained median values are 0.30 for the anechoic case and 0.69 for reverberation and indoor scenarios. Although these two latter take the same median value, slight differences are observed between their distributions. For $\Delta$ values lower than the median, the distribution for the indoor configurations is higher than for the reverberation ones. In contrast, for higher $\Delta$ values, the roles are exchanged and the distribution becomes lower for the indoor case. This is explained by the mixed character of these types of scenarios. The studied indoor measurements combine the LoS situation with a rich multipath. The high power of the LoS rays produces an imbalance between the amplitudes of the received rays which implies a reduction of the $\Delta$ parameter.

The $m_1$ and $m_2$ fluctuation parameters distributions are depicted in Figs. 4(c) and 4(d). The first ray fluctuates more in the reverberation and indoor scenarios than in the anechoic ones since the $m_1$ parameter is distributed along low values (the medians are 1 and 28 respectively). That is due to the multipath effect, which provides more accentuated variations in the received signal. This also explains that distribution for reverberation and indoor are overlapped between each other. On the other hand, the $m_2$ parameter takes higher values than the $m_1$ in the multipath scenarios (median values of 20 and 21 respectively). This implies that, generally, the second ray is less fluctuating than the first for the reverberation and indoor cases, whereas for the anechoic both rays have reduced fluctuation. Nevertheless, for the indoor configurations, there exists a remarkable percentage of channels with a low $m_2$ parameter which is related to a significant multipath effect.

Some particular fitting examples\footnote{As supplementary material for the reader, an additional animation is included showing the obtained fitting solutions for each of the 142 configuration{s} in the three studied scenarios (see Appendix B).} configurations (namely as A, B, C, D, E, and F) are depicted in Fig. \ref{fig:fitting_solutions}. A good fitting accuracy is observed for the three scenarios, and the particular numerical results can be consulted in Table~\ref{tab:fitting_results}. Bimodal behaviors are accurately modeled, as well as \linebreak Rician-like and Rayleigh-like ones common in multipath scenarios. Specifically, the IFTR model is capable of reproducing bimodality as long as $K$ takes large values and $\Delta$ is near to one. These conditions are equivalent to those explained in Section III for experimental bimodality. That is, the power of the first ray has to be comparable to the one of the second ray, and both have to be notably higher than the diffuse one. Finally, Fig.~\ref{fig:configurations} shows the specific configurations of the given examples in terms of the two combined Rx array positions (points with the same color) jointly with the common Tx orientation. Positions are referred with respect to the array center, which corresponds to the (0,0) coordinate.

The IFTR model is able to properly describe the expected characteristics from such diverse scenarios (anechoic, reverberation and indoor environments) under very different conditions. Since from the set of merged channels multiple behaviors emerge, this ray-based model exhibits a great versatility for describing complex scenarios. As mentioned in the previous section, these merged channels may resemble a great variety of indoor scenarios (e.g. offices, meeting rooms...). Thus, the model applicability covers multitude of different channels and environments.

\subsection{Anechoic environment analysis}
Although most of the anechoic configurations can be described with high accuracy, there exists a certain percentage of fitted configurations with an unsatisfactory \ac{GoF}. After a thorough analysis, the key aspect that justifies this behavior is the lack of phase diversity in certain combinations. That is, the traveled path by the waves and the deep attenuation of diffuse components is insufficient to produce an uniform distribution of phase differences between $-\pi$ and $\pi$. This has two main implications: on the one hand, it causes the \ac{CLT} assumption for the diffuse component is not met; on the other hand, it affects the assumption of uniformly distributed phases for the dominant specular components. The latter is of importance since the received amplitude of the IFTR amplitude can be formulated in terms of the phase difference $\alpha$ as follows:
\begin{equation}
    V_r = \exp\left(j\varphi_1\right)(A_1 + A_2\exp\left(j\alpha\right)) + Z,
\end{equation}
where $\alpha=\varphi_2 - \varphi_1$ follows an uniform distribution in the range from $-\pi$ to $\pi$.

\begin{figure}[t]
    \centering
    \includegraphics{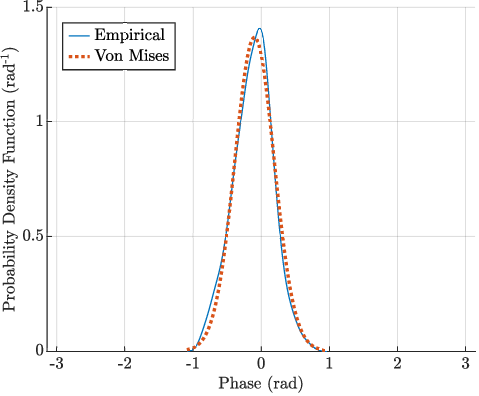}
    \caption{Phase difference distribution for the anechoic configuration G. The dashed line represents the fit with the von Mises PDF.}
    \label{fig:non-uniform_distribution}
\end{figure}

\begin{table}[t]
    \centering
    \renewcommand{\arraystretch}{1.75}
    \caption{{Numerical results} comparison \linebreak between the \ac{IFTR} and the GTR-V models}
    \begin{tabular}{c|c|ccc|}
        \multicolumn{1}{c}{} & \multicolumn{1}{c}{} & \multicolumn{3}{c}{\textbf{Anechoic configuration}} \\ \cline{3-5}
        \multicolumn{1}{c}{\textbf{Model}} & \multicolumn{1}{c|}{\textbf{Parameter}} & \textbf{G} & \textbf{H} & \textbf{I} \\ \hline
        \multirow{4}{*}{\ac{IFTR}} & $K$ [\si{\decibel}] & \num{30.0} & \num{30.0} & \num{25.1} \\
        & $\Delta$ & \num{2.7e-3} & \num{0.7e-3} & \num{0.40} \\
        & $(m_1,m_2)$ & $(\num{50},\num{32})$ & $(\num{50},\num{5})$ & $(\num{31},\num{13})$ \\ \cdashline{2-5}
        & $\mathrm{RMSE}$ & \num{142.38} & \num{28.17} & \num{17.58} \\ \hline
        \multirow{6}{*}{GTR-V} & $K$ [\si{\decibel}] & \num{19.8} & \num{23.8} & \num{10.6} \\
        & $\Delta$ & \num{0.45} & \num{2.6e-4} & \num{0.11} \\
        & $\phi$ & \num{-0.10} & \num{0.08} & \num{-0.10} \\
        & $\kappa$ & \num{12.04} & \num{638.94} & \num{1.23}  \\ \cdashline{2-5}
        & $\mathrm{MSE_{VM}}$ & \num{2.9e-3} & \num{0.19} & \num{6.2e-4} \\
        & $\mathrm{RMSE}$ & \num{37.66} & \num{10.04} & \num{28.66} \\ \hline
    \end{tabular}
    \label{tab:GTR-V_fit_solution_data}
\end{table}

\begin{figure*}[t!]
    \centering
    \subfigure[Configuration G.]{
        \centering
        \includegraphics{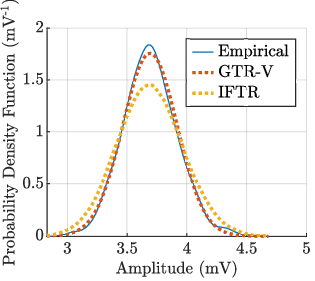}
        \label{fig:gtrv-g}
    }
    \hspace{0.015\linewidth} 
    \subfigure[Configuration H.]{
        \centering
        \includegraphics{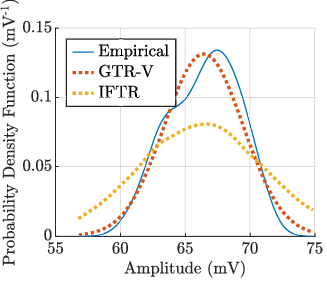}
        \label{fig:gtrv-h}
   }
   \hspace{0.015\linewidth} 
    \subfigure[Configuration I.]{
        \centering
        \includegraphics{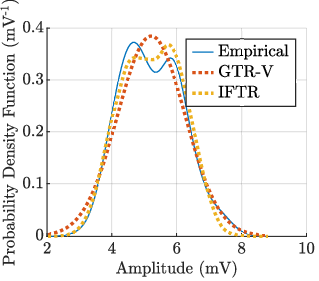}
        \label{fig:gtrv-i}
    }
    \caption{Comparison between fitted \ac{IFTR} and \ac{GTR-V} \acp{PDF} for several anechoic configurations.}
    \label{fig:GTR-V_fit_solutions}
\end{figure*}

The implications of a non-uniform phase distribution have been discussed in \cite{Rao2015_GTR-V}, where the \ac{TWDP} model was modified to account for this particular situation. The resulting model, termed as \ac{GTR-V}, has the following \ac{PDF} for the received amplitude $r$:
\begin{equation}\label{eq:GTR-V PDF}
    f_{\rm GTR-V}(r) = \int_{-\pi}^{\pi} f_{\rm rice}(r;K[1 + \Delta\cos(\alpha)])f_{\alpha}(\alpha)\mathrm{d}\alpha,
\end{equation}
\noindent where $f_{\rm rice}({r,K_r})$ is the well-known Rician distribution with a factor $K_r = K[1 + \Delta\cos(\alpha)]$. The PDF is given by
\begin{equation}\label{eq:Rician PDF}
    f_{\rm rice}(r;K_r) = \frac{r}{\sigma^2}\exp\left(-\frac{r^2}{2\sigma^2} - K_r\right)I_0\left(\frac{r}{\sigma}\sqrt{2K_r}\right),
\end{equation}
\noindent with $I_0(x)$ being the modified Bessel function of first class and order zero. On the other hand, $f_{\alpha}(\alpha)$ represents the phase difference distribution. In the \ac{GTR-V} channel model, it corresponds to a von Mises \ac{PDF} \cite{von_mises_distribution}:
\begin{equation}\label{eq:von Mises PDF}
    f_\alpha(\alpha;\kappa,\phi) = \frac{\exp(\kappa\cos(\alpha - \phi))}{2\pi I_0(\kappa)} \:\: \forall \:\: \alpha \in [-\pi,\pi],
\end{equation}
\noindent where $\phi \in \mathbb{R}$ represents the mean of the distribution and $\kappa \geq 0$ is inversely related to its variance. 

Indeed, it is possible to generalize the IFTR model to incorporate the effect of a non-uniform phase distribution following the reasoning of \cite{Rao2015_GTR-V}. Nevertheless, this would imply that the resulting model would have two additional parameters, incurring additional complexity when it comes to fitting. Instead, we will use the \ac{GTR-V} model even though it has no fluctuation on the two dominant specular components; in practice, this is equivalent to assuming a sufficiently large value of $m_1$ and $m_2$ in the \ac{IFTR} model (see Fig. \ref{fig:iftr_parameters_cdf}). Therefore, the use of this alternative model is well-justified for the sake of simplicity. After analyzing the distribution of the phase difference between the dominant specular components in the relevant anechoic configurations, a small sub-set with von Mises-like \ac{PDF} for phase difference has been selected (named configurations G, H, and I).

The fitting process for the \ac{GTR-V} model has been carried out as a two-step optimization problem. First, the empirical phase difference distribution is modeled as a von Mises distribution, finding the two parameters $\phi$ and $\kappa$ that provide the best fit for the phase difference distribution. After finding these parameters, a second optimization is performed analogously as for the IFTR model, finding the best pair $\{K,\Delta\}$ for the \ac{GTR-V} model applying a \ac{GA} search.

\begin{figure}[p!]
    \centering
    \subfigure[Configuration J (reverberation).]{
        \centering
        \includegraphics{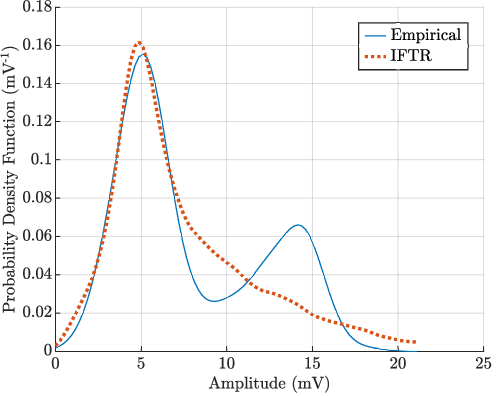}
    }
    \par\bigskip
    \subfigure[Configuration K (indoor).]{
        \centering
        \includegraphics{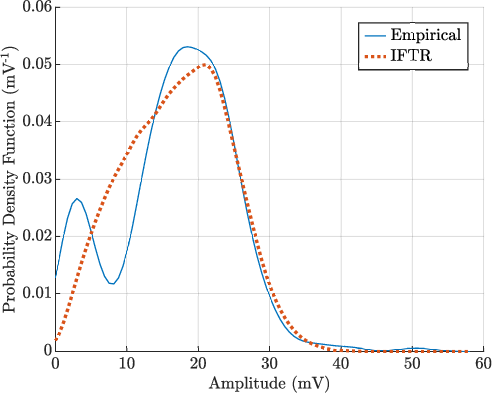}
    }
    \caption{Empirical and fitted \acp{PDF} for sharp bimodality behaviors.}
    \label{fig:bad_fitting_multipath}
\end{figure}

Fig. \ref{fig:non-uniform_distribution} shows the fitting of the phase difference distribution\footnote{Though the channel measurements include both diffuse and specular components, in anechoic channels the LoS ray power is significantly the highest. Thus, it is reasonable to estimate the phase difference distribution as the PDF of the phase difference of the two merged channels.} with a von Mises \ac{PDF} for configuration G. After the first step of the optimization, the resulting parameters $\kappa$ and $\phi$ yield an MSE of 0.0029, which implies an excellent \ac{GoF}. This process has been carried out also for H and I configurations. The next step is to find the optimal $K$ and $\Delta$ parameters for the GTR-V model by the use of \ac{GA}. The obtained results for the configurations G, H, and I are summarized in Table~\ref{tab:GTR-V_fit_solution_data}, comparing them with the obtained for the IFTR model. In addition, the obtained \acp{PDF} are depicted in Fig. \ref{fig:GTR-V_fit_solutions}. For configurations G and~H [Figs. \ref{fig:gtrv-g} and \ref{fig:gtrv-h}], it is observed that the \ac{GTR-V} model provides better results than the IFTR model. The effect of a non-uniform phase distribution is translated into an effective reduction of the variance, and it can be visually confirmed that the \ac{GTR-V} model achieves this. However, we see that the fitting is not improved when the empirical \ac{PDF} is bimodal --- see configuration I in Fig. \ref{fig:gtrv-i}. Thus, the capabilities of the TWDP model to capture bimodality are affected due to the non-uniform phase difference distribution. 

\begin{figure}[t]
    \centering
    \subfigure[Configuration J (reverberation).]{
        \centering
        \includegraphics{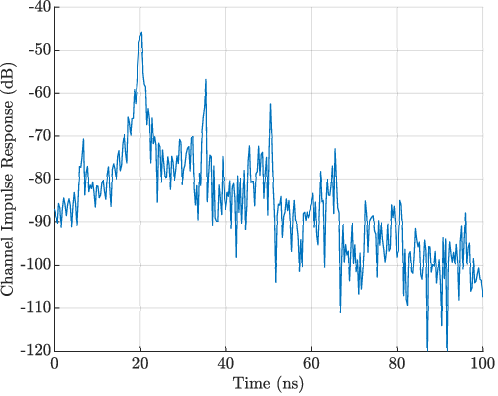}
        \label{fig:cir-j}
    }
    \par\bigskip
    \subfigure[Configuration K (indoor).]{
        \centering
        \includegraphics{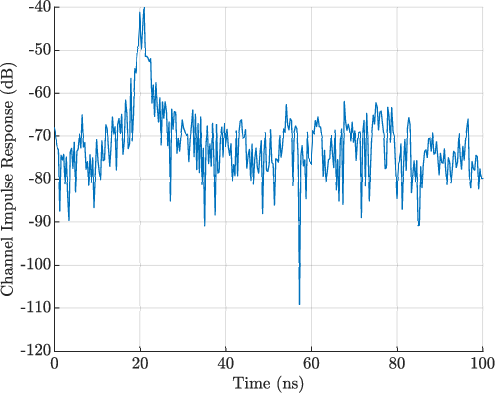}
        \label{fig:cir-k}
    }
    \caption{\acp{CIR} for multipath configurations with sharp bimodality behaviors.}
    \label{fig:cir_multipath}
\end{figure}

\subsection{Multipath environments analysis}
For the multipath cases, most of the configurations are described correctly with the \ac{IFTR} model. However, we have identified some cases where it fails. Specifically, in situations with sharp bimodal behaviors, the ability of the \ac{IFTR} model to capture such bimodality is not enough. This is exemplified in Fig.~\ref{fig:bad_fitting_multipath} for another configurations J and K in reverberation and indoor environments respectively. In these situations, the model tries to fit the dominant peak but masks the secondary one. Numerical results can be consulted in Table~\ref{tab:multipath_bad}.

\begin{table}[t!]
    \centering
    \renewcommand{\arraystretch}{1.75}
    \caption{Numerical results of the fitting between the IFTR model and the multipath channels with sharp bimodality behaviors}
    \begin{tabular}{c|cc|}
        \multicolumn{1}{c}{} & \multicolumn{2}{c}{\textbf{Multipath configuration}} \\ \cline{2-3}
        \multicolumn{1}{c|}{\textbf{Parameter}} & \textbf{J (reverberation)} & \textbf{K (indoor)} \\ \hline
        $K$ [\si{\decibel}] & \num{28.9} & \num{28.2} \\
        $\Delta$ & \num{0.92} & \num{0.87} \\
        $(m_1,m_2)$ & $(\num{0.45},\num{30})$ & $(\num{3},\num{31})$ \\ \cdashline{1-3}
        $\mathrm{RMSE}$ & \num{18.27} & \num{6.33} \\\hline
    \end{tabular}
    \label{tab:multipath_bad}
\end{table}

One way to determine the origin of these particular behaviors is to analyze the \ac{CIR} module of the configurations (see Fig. \ref{fig:cir_multipath}). On the one hand, Fig. \ref{fig:cir-j} depicts the CIR for the reverberation configuration~J. It is clear that more than two rays are arriving at the antenna with power significantly high and comparable between each other. This is not covered by the \ac{IFTR} model. Even though the consideration of additional rays can be usually encompassed by integrating these into the diffuse component \cite{Romero2022}, this is not the case when the overall number of rays is reduced, multipath propagation is limited, and there is a lack of phase richness in the propagation environment. On the other hand, Fig. \ref{fig:cir-k} presents the CIR for the indoor configuration K. Two dominant rays are observed, but they are overlapped in time between each other. Hence, its interaction may not be described independently as the IFTR model assumes. The sharp bimodality behavior appears as a cause of the time interaction of these two rays.

\section{Conclusions}\label{sec:Conclusions}
We conducted an empirical validation of the independent fluctuating two-ray (IFTR) fading model, covering a wide range of scenarios in the frequency band n258 (from 24.25\:GHz to 27.5\:GHz). We confirmed that the IFTR model is highly versatile, and is able to recreate rather dissimilar propagation conditions including anechoic, reverberation, and indoor scenarios. Besides, the four shape parameters that characterize the model have a solid physical meaning, which agrees with the expected properties of each analyzed channel. That is, these few parameters are sufficient to successfully capture the particularities of propagation environments without the need to resort to more complex models with a larger number of parameters.

We also observed some relevant effects that put forth the limitations of the IFTR model: (\textit{i}) the assumption of uniform phases for the dominant specular components does not always hold, especially when there is a lack of phase richness. This is the case in scenarios with a reduced multipath, and is likely to become a dominant effect as we move up in frequency; (\textit{ii}) the IFTR model is not able to capture sharp bimodal behaviors observed in some of the measurements. This can be caused by the combination of a number of effects, such as the presence of additional rays (due to spurious reflections), the absence of rich multipath propagation, and the interaction between dominant specular components. In this regard, the development of physically-motivated models that capture these behaviors deserves special attention, together with additional empirical validations at higher and lower frequencies; and also in new different scenarios. 

\section*{Appendix A \\ IFTR Channel Model Fitting}
Although moment-based or maximum-likelihood (ML) estimators are often used with sample data to fit a model distribution, cases where the distribution has a complicated form often require to be solved numerically \cite{ref:Talukdar1991}. Alternatively, the empirical PDF can be estimated and then, an optimization problem is solved to obtain the values of the model parameters that minimize an error metric. In our case, to perform the fit between the \ac{IFTR} model and our measurements we have applied a methodology based on \ac{GA}\footnote{Other alternatives such as the MATLAB in-built routines fminsearch and lsqcurvefit were also implemented \cite{ref:Anjos2024}, obtaining a comparable but slightly worse performance compared to the GA method \cite{ref:Gomes2024}.}, specifically the so-called non-dominated sorting genetic \linebreak algorithm II (NSGA-II)~\cite{Deb2001_multiobjective_optim}.

A multi-objective optimization problem has been defined with the following set of error metrics\footnote{The $f_{\rm exp}(x)$ and $f_{\rm mod}(x)$ functions represent the empirical and the \ac{IFTR} channel model \acp{PDF} respectively. The parameter $N$\:=\:100, and corresponds with the number of \ac{PDF} sampled points.} as targets:
\begin{itemize}
    \item Mean Squared Error (MSE):
    \begin{equation}\label{eq:mse_error}
        \mathrm{MSE} \triangleq \frac{1}{N}\sum_{i=1}^{N}(f_{\rm exp}(x_i)-f_{\rm mod}(x_i))^2.
    \end{equation}
    \item Root Mean Squared Error (RMSE):
    \begin{equation}\label{eq:rmse_error}
        \mathrm{RMSE} \triangleq \sqrt{\frac{1}{N}\sum_{i=1}^{N}(f_{\rm exp}(x_i)-f_{\rm mod}(x_i))^2}.
    \end{equation}
    \item Mean Absolute Error (MAE):
    \begin{equation}\label{eq:mae_error}
        \mathrm{MAE} \triangleq \frac{1}{N}\sum_{i=1}^{N}|f_{\rm exp}(x_i)-f_{\rm mod}(x_i)|.
    \end{equation}
    \item Modified Kolmogorov-Smirnov (KS) statistic:
     \begin{equation}\label{eq:ks_error}
        \mathrm{KS} \triangleq \max_{x_i}|f_{\rm exp}(x_i)-f_{\rm mod}(x_i)|.
    \end{equation}
\end{itemize}
The use of multiple objective functions provides the \ac{GA} of a greater robustness to search for optimal solutions. Additionally, it aims to find differences between the results when minimizing each error metric.

As an evolutionary algorithm, NSGA-II requires both a population size and a maximum number of generations to be defined. In our specific case, since we only need to tune four parameters, a population of 200 individuals and 400 generations is adequate. Another relevant parameter is the elite count (number of individuals guaranteed to survive to the next generation), which has been established to the 5\% of the population size. Finally, as stopping condition, we have configured the algorithm to stop when 100 generations occur consecutively without significant changes in the solutions.

\begin{figure}[t]
    \centering
    \includegraphics{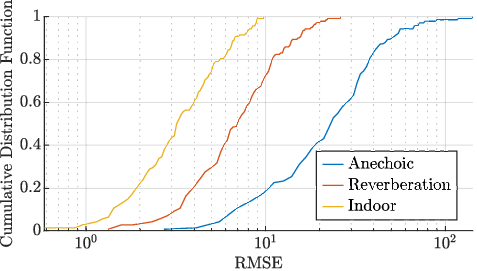}
    \caption{RMSE metric CDFs for the optimal IFTR solutions.}
    \label{fig:errors_cdf}
\end{figure}

\begin{figure}[t]
    \centering
    \includegraphics{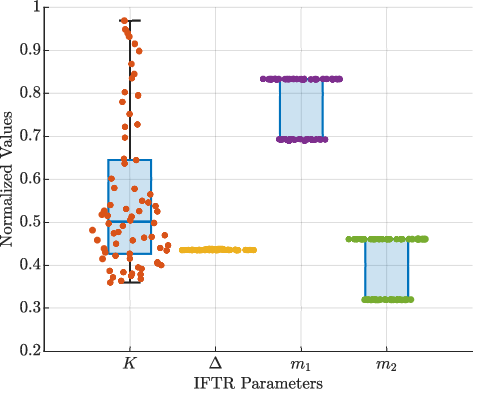}
    \caption{{Boxplot diagrams of the} \ac{IFTR} parameters for the Pareto front obtained for the anechoic configuration A. The individual points represent the solutions for each parameter, whereas the boxplots show their distribution. Values are normalized with respect to the upper limits established for the \ac{GA} optimization search (that is, $K$ w.r.t.~10\textsuperscript{3}; and $m_1$ and $m_2$ w.r.t. 30).}
    \label{fig:pareto_front_two_sols}
\end{figure}

Another relevant concept in the optimization problem is the solutions space, which is infinite. Thus, it is necessary to define upper and lower bounds for limiting the search along each variable. For the $K$-factor, we have restricted the search to the interval from 0 to 10\textsuperscript{3} (that is, up to 30\:dB). On the other hand, for the $m_1$ and $m_2$ parameters we have set 0 and~50 as upper and lower limits respectively. Finally, since the $\Delta$ parameter takes values in the interval [0,1], the search has been bounded to the complete definition domain. As can be seen in Fig. \ref{fig:iftr_parameters_cdf}, there are no discontinuities on the distributions at the upper/lower boundary values. For instance, if it was not the case, it would mean that the GA saturated into the limit value of 30\:dB in the case of the $K$-factor. Therefore, the established bounds for the GA search are adequate providing a good GoF as Fig. 5 shows.

\begin{figure}[t]
    \centering
    \includegraphics{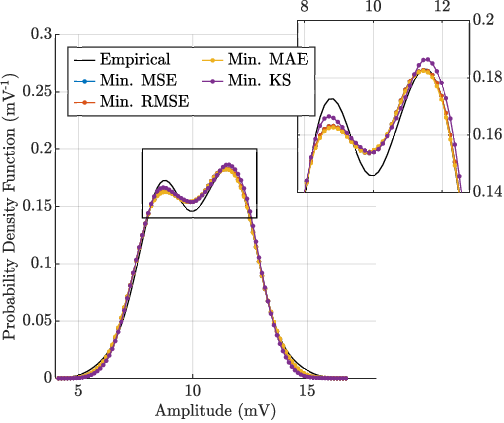}
    \caption{{\ac{IFTR} \ac{PDF} fitting results obtained for the anechoic configuration A minimizing each error metric. The empirical \ac{PDF} is included as a reference.}} 
    \label{fig:pareto_comparison}
\end{figure}

For each fitted channel, the output of the \ac{GA} is a set of~$N_p$ solutions known as Pareto front. Each solution consists of a tuple $(K,\Delta,m_1,m_2)$ representing a compromise situation between the four objective functions. To choose a particular solution $\hat{S}$\:=\:$(\hat{K},\hat{\Delta},\hat{m}_1,\hat{m}_2)$ from the Pareto front, we need some unified criterion. For that purpose, we first define the normalized error functions. As an example, the normalized $\mathrm{MSE}_n$ function would be
\begin{equation}\label{eq:mse_normalized_error}
    \mathrm{MSE}_n(S) \triangleq \frac{\mathrm{MSE}(S)}{\displaystyle \max_{P}\mathrm{MSE}(P)}.
\end{equation}
The normalization factor represents the maximum error metric obtained from among all the solutions of the Pareto front. Definition of $\mathrm{RMSE}_n$, $\mathrm{MAE}_n$ and $\mathrm{KS}_n$ functions is completely analogous. From them, we can define a normalized metric $\varepsilon_n$ as follows:
\begin{equation}\label{eq:mean_normalized_error_metric}
    \varepsilon_n({S}) \triangleq \left.\frac{\mathrm{MSE}_n + \mathrm{RMSE}_n + \mathrm{MAE}_n + \mathrm{KS}_n}{4}\right|_{{S}}.
\end{equation}
Thus, selected solution $\hat{S}$ will correspond with the $S$ that minimize the metric $\varepsilon_n$, that is
\begin{equation}
    \hat{{S}} = \min_{{S}} \varepsilon_n({S}).
\end{equation}
This equally-weighted normalized metric has been selected for the sake of simplicity. Since all the solutions from the Pareto front result in similar absolute error metrics, the choice of weights was not a key factor during our experiments. We give the distribution of the RMSE for the selected solutions as a reference for the fitting accuracy (see Fig. \ref{fig:errors_cdf}). The lowest error is obtained for the indoor case, with a value of 3.2; whereas for the reverberation and anechoic cases the medians are 7.0 and 23.0 respectively.

\begin{table}[t]
    \centering
    \renewcommand{\arraystretch}{1.75}
    \caption{\ac{IFTR} parameters for each minimum error metric solution of the Pareto front in anechoic configuration A}
    \begin{tabular}{c|cccc|}
        \multicolumn{1}{c}{} & \multicolumn{4}{c}{\textbf{Min. Error Metric Solutions (Conf. A)}} \\ \cline{2-5}
        \multicolumn{1}{c|}{\textbf{Parameter}} & \textbf{MSE} & \textbf{RMSE} & \textbf{MAE} & \textbf{KS} \\ \hline
        Min. error & \num{16.56} & \num{4.07} & \num{2.96} & \num{8.18} \\ \hline
        $K$ [\si{\decibel}] & \num{28.6} & \num{28.6} & \num{27.2} & \num{27.3} \\
        $\Delta$ & \num{0.44} & \num{0.44} & \num{0.44} & \num{0.44} \\
        $(m_1,m_2)$ & $(\num{35},\num{16})$ & $(\num{35},\num{16})$ & $(\num{35},\num{16})$ & $(\num{42},\num{23})$ \\ \hline
    \end{tabular}
    \label{tab:pareto_comparison_results}
\end{table}

Although one solution from the complete Pareto front has to be selected, it is interesting to analyze the effects of the individual error functions on the fits. For instance, Fig. \ref{fig:pareto_front_two_sols} shows the solutions obtained for the Pareto front of configuration A (see Fig. \ref{fig:complete_channel_scheme_in_reverse_path}) in the anechoic case. Boxplot diagrams and swarm-chart scatter plots are represented jointly for each parameter (normalized with respect to its upper bound). The $K$-factor is distributed along a large interval which ranges from 0.36 to 0.97 (or equivalently from 25.6\:dB to 29.8\:dB, units no-normalized). However, most of the solutions are distributed near the median value of 0.50 (27.0\:dB). In contrast, the $\Delta$ parameter practically does not change taking the same value of 0.44 for all the solutions tuples. Differences appear for the fluctuation parameters $m_1$ and $m_2$, emerging two well-defined values. Specifically, for the parameter $m_1$ are obtained 0.69 and 0.83 (34.5 and 41.5 in no-normalized units); meanwhile for the $m_2$ are obtained 0.32 and 0.46 (16 and 23 without normalizing). This particular behavior is common for the majority of the studied merged channel configurations in the three scenarios. We can focus our attention on the solutions tied to the minimum of each error metric (see Table \ref{tab:pareto_comparison_results}). The two values for the pair $m_1$ and $m_2$ are either associated to the average error metrics (MSE, RMS and MAE) or to the distance error metric~(KS) respectively. As previously mentioned, $\Delta$ remains almost unchanged in the four solutions tuples. Finally, the $K$-factor is equal for the MSE and the RMSE solutions and differs roughly 1\:dB with those obtained for the MAE and KS solutions. In Fig.~\ref{fig:pareto_comparison} are shown the experimental \ac{PDF} of the anechoic configuration A and the theoretical \ac{IFTR} curves for each of the mentioned solutions, which are practically overlapped between each other. On the one hand, the \acp{PDF} associated with the MSE, RMSE and MAE error metrics have the same shape (quasi-equal IFTR parameters). On the other hand, the curve of the minimum KS is slightly different improving the fitting on the modal regions at the expense of the distribution tails. The latter are fitted with more accuracy by minimizing the mean errors.

\section*{Appendix B \\ Supplementary Material}
The supplementary material is available at the IEEE DataPort repository: https://dx.doi.org/10.21227/hzw9-6q21. It consists of three different animations, where two of them present the channel diversity along the computed combinations and the other one the obtained solutions for the \ac{IFTR} fitting. The former shows multitude of configurations with a representative scheme, the associated \ac{PDF} of received amplitude for each scenario, and the phase difference \ac{PDF}. The latter represents the obtained solutions for the Pareto front, the distribution of the \ac{IFTR} parameters and the RMSE metric, the phase \ac{PDF}, and the associated configuration scheme.

\end{document}